\documentclass[aip,pop,amsfonts,amsmath,amssymb,table,preprint]{revtex4-1}
\pdfoutput=1 

\usepackage{graphicx}

\usepackage[english]{babel}
\usepackage[utf8]{inputenc}
\usepackage[pdftitle={Article}, pdfauthor={Author}]{hyperref} 

\begin{document}


\title{Influence of finite ion Larmor radius on the dynamics of weakly-collisional plasma jets colliding in magnetic arch} 

\author{Artem V. Korzhimanov}
\email[Correspondence email address: ]{artem.korzhimanov@ipfran.ru}
\affiliation{Federal Research Center A. V. Gaponov--Grekhov Institute of Applied Physics of the Russian Academy of Sciences, Nizhny Novgorod, Russia}
\affiliation{Lobachevsky State University of Nizhny Novgorod, Nizhny Novgorod, Russia}
\author{Roman S. Zemskov}
\affiliation{Federal Research Center A. V. Gaponov--Grekhov Institute of Applied Physics of the Russian Academy of Sciences, Nizhny Novgorod, Russia}
\author{Sergey A. Koryagin}
\author{Mikhail E. Viktorov}
\affiliation{Federal Research Center A. V. Gaponov--Grekhov Institute of Applied Physics of the Russian Academy of Sciences, Nizhny Novgorod, Russia}
\affiliation{Lobachevsky State University of Nizhny Novgorod, Nizhny Novgorod, Russia}

\date{\today}

\begin{abstract}
The effect of the finite ion Larmor radius on the dynamics of two counterstreaming weakly collisional plasma flows in a magnetic field of an arch configuration is considered. Hybrid numerical simulations show that in a system whose dimensions are close to the ion Larmor radius, more intense interaction dynamics are observed, the magnetic arch experiences a significant expansion with the formation of a region with an irregular character of magnetic lines, in which magnetic reconnection processes occur. In this case, the generation of a surface wave of the ion-cyclotron range is observed at the boundaries of the arch. An increase in the scale of the system compared to the ion Larmor radius leads to a transition to the ideal MHD regime, in which the evolution of the arch occurs much more slowly, and the development of instabilities is not observed.
\end{abstract}

\pacs{}

\maketitle 

\section{\label{sec:intro}Introduction}
The interaction of plasma flows with a magnetic field of an arched configuration is widely encountered both in astrophysics, for example, in solar flares \citep{masuda_N_1994, xia_A_2020, guo_RMPP_2024} and in the magnetosphere of planets \citep{liu_AAST_2020}, and in technical devices, for example, in thermonuclear reactors based on magnetic plasma confinement \citep{kikuchi_RMP_2012}. At the same time, the study of such systems faces a number of technical difficulties that limit the possibilities for diagnosing plasma and fields with the time and spatial resolution necessary for understanding the processes taking place. This makes it valuable to study model processes in simplified laboratory conditions, in which controlled and reproducible generation of plasma flows in given magnetic fields is possible.

Recently, a number of studies have been conducted on the interaction of plasma flows generated in the bases of a magnetic arch \citep{katz_PRL_2010, stenson_PRL_2012, tripathi_PRL_2010, zhang_NA_2023}. In particular, our group has developed a laboratory setup based on arc discharge plasma injected into a field generated by a pair of pulse coils located at an angle to each other \citep{viktorov_TPL_2015, viktorov_PPCF_2019}. The discharge time is 20 $\mu$s, which is enough to fill the volume of a chamber with a diameter of 20 cm during the discharge. Under typical conditions, the setup is capable of creating magnetic fields of the order of $B_0=10$--$100$ mT in the center of the chamber and plasma flows moving at a velocity of $V_0\approx 10^6$ cm/s, with a particle concentration in the range of $N_0 = 10^{13}$--$10^{16}$ cm$^{-3}$. The electron and ion temperature $T_0$ in each flow does not exceed several eV, and the energy of directed ion motion $W_{i0} = M_iV_0^2/2$ ($M_i$ is the ion mass) is several times higher, for example, for aluminum plasma moving at a velocity of $10^6$ cm/s, $W_{i0}\approx 14$ eV. Thus, in this setup it is possible to observe both sub-Alfven and super-Alfven flows: for example, for the same plasma at $B_0=100$ mT the magnetic Mach number $M_m = V_0/V_A$ ($V_A = B_0/\sqrt{2\mu_0N_0M_i}$ is the Alfven velocity, $\mu_0$ is the magnetic constant) is equal to unity at $N_0 = 1.76\times 10^{15}$ cm$^{-3}$.

We have previously shown that the transition from the sub-Alfvenic regime to the super-Alfvenic one leads to a significant change in the dynamics of plasma flows in such a system \citep{korzhimanov_PP_2025}. In the sub-Alfvenic regime, a more or less stable plasma arch is formed, slowly evolving mainly due to the $\mathbf E\times \mathbf B$ drift with the formation of a region of oppositely directed magnetic field lines, in which, however, no intense magnetic reconnection is observed. In the super-Alfvenic regime, a partial breakthrough of magnetic lines by plasma flows occurs and a region of turbulent plasma is formed, in which magnetic reconnection processes are more intense and the formation of plasmoids is observed.

Estimates show that in the regime when $M_m\sim 1$, electrons are collisional with a characteristic collision frequency of $10^{10}$ s$^{-1}$ and a gyrofrequency of the order of $10^{9}$ s$^{-1}$, while ions are weakly collisional with a characteristic collision frequency of $10^{4}$ s$^{-1}$ and a gyrofrequency of the same order. Thus, the system is subject to the development of ion kinetic instabilities. In particular, the excitation of surface ion-cyclotron waves arising due to the anisotropy of the ion distribution function was observed in the simulation.

This system, however, features the relatively small size of the generated plasma flows. The diameter of the outgoing hole from which the plasma flows out is 2 cm, which is less than or of the same order as the Larmor radius of the ions. Thus, the effects of the finite ion Larmor radius (FILR) become significant for the dynamics of the interaction. Note that the effect of FILR on the processes of interaction of plasma flows with a magnetic field has been the subject of research for many years. In particular, its influence on various instabilities developing in plasma was discussed, for example, drift-cyclotron instability in inhomogeneous plasma \citep{mikhailovsky_NF_1965}, Rayleigh-Taylor instabilities during plasma expansion in a magnetic field \citep{huba_PRL_1987, ripin_PRL_1987, tang_PP_2020b} and others \citep{pokhotelov_JGR_2004, ferraro_A_2007, landreman_PRL_2015}. The influence of FILR on the type and properties of waves near the ion cyclotron frequency \citep{brambilla_PPCF_1989, kolesnichenko_JPP_2023, kolesnichenko_PP_2024} was noted. FILR is also of great importance in the theory of collisionless magnetic reconnection \citep{grasso_PPR_2000, sarto_PPCF_2011} and play a significant role in the magnetosphere \citep{stasiewicz_SSR_1993}, the solar atmosphere \citep{pandey_MNRAS_2022} and in the interplanetary plasma \citep{kubo_A_2010}.

The aim of this work is a numerical study of the influence of FILR on the processes occurring in the experimental setup in the regime of magnetic Mach numbers of the order of unity.

\section{\label{sec:methods}Numerical methods}
For collisional electrons and weakly collisional ions, the optimal modeling method is a hybrid one, in which ions are described in the kinetic collisionless approximation by the particle-in-cell method, and electrons are described as a massless neutralizing liquid. To partially take into account the relatively weak kinetic electron effects, electrons are described in the so-called 10-moment approximation, taking into account the pressure tensor evolution equation. Due to the low flow velocities, the electromagnetic field is described in the low-frequency (Darwinian) approximation, in which the displacement current is neglected. Thus, the complete system of equations to be solved takes the form \citep{Hesse1995}:
\begin{gather}
    \frac{\partial f_i}{\partial t} + \mathbf v_i\frac{\partial f_i}{\partial \mathbf r} + \frac{Ze}{M_i}\left(\mathbf E + \mathbf v_i\times\mathbf B\right)\frac{\partial f_i}{\partial \mathbf v_i} = 0 \\
    n_e = Zn_i = Z\int f_i(t,\mathbf r,\mathbf v_i)\mathrm d \mathbf v_i \\
    \mathbf V_i = \frac{1}{n_i}\int \mathbf v_if_i(t,\mathbf r,\mathbf v_i)\mathrm d \mathbf v_i \\
    \mathbf E = -\mathbf V_i\times\mathbf B + \frac{1}{en_e}\left(\mathbf j\times\mathbf B - \nabla.\mathbb{P} \right) \label{eq:ohm}\\
    \frac{\partial \mathbf B}{\partial t} = -\mathrm{rot}\,\mathbf E \\
    \mathbf j = \frac{1}{\mu_0}\mathrm{rot}\,\mathbf B \\
    \mathbf V_e = -\frac{1}{en_e} \mathbf j + \mathbf{V_i} \\
    \frac{\partial \mathbb{P}}{\partial t} + \mathbf V_e.\nabla\mathbb{P} = - \mathbb{P}\nabla.\mathbf V_e - \mathbb{P}.\nabla\mathbf V_e - \left(\mathbb{P}.\nabla\mathbf V_e\right)^T - \frac{e}{m_e}\left[\mathbb{P}\times\mathbf B + \left(\mathbb{P}\times\mathbf B\right)^T\right]
\end{gather}
Here $e$, $m_e$ are the elementary charge and mass of an electron, $Z$ is the ionization multiplicity of ions, $n_{e,i}(t,\mathbf r)$ are the concentrations of electrons and ions, respectively, $\mathbf V_{e,i}(t,\mathbf r)$ are the average (hydrodynamic) velocities of electrons and ions, $\mathbb{P}$ is the electron pressure tensor, $\nabla$ is the del (nabla) vector differential operator, and $(\cdot)^T$ denotes the tensor transpose. Note that the kinetic description of the ion dynamics explicitly takes into account the finiteness of their Larmor radius, while its influence on the plasma dynamics as a whole is provided by the second and third terms in the generalized Ohm's law (\ref{eq:ohm}), while the term $\mathbf j\times\mathbf B$, called the Hall term, is responsible for the effects associated with the directed velocity of ions relative to electrons, and the term $\nabla.\mathbb{P}$ is responsible for the effects associated with their thermal velocities. These terms are responsible for the effects of non-ideal magnetohydrodynamics. In the generalized Ohm equation, we neglected the term associated with the electron inertia, which is justified in the case of a sufficiently hot plasma, when the plasma beta $\beta > m_e/M_i$ (which is equivalent to the smallness of the electron inertial length compared to the ion Larmor radius), which is certainly satisfied under the conditions of our setup.

Hybrid modeling was performed using the AKA code \citep{sladkov_JPCS_2020}. It was tested in a model problem for analyzing the magnetic reconnection process in the Harris layer on electron spatial scales \citep{sladkov_PP_2021}, in a problem on the long-term dynamics of the Weibel instability \citep{sladkov_2023}, as well as in modeling experiments on laser ablation of plasma in an external magnetic field \citep{burdonov_A_2022, bolanos_NC_2022, zemskov_A_2024, sladkov_NC_2024}.

The simulation was close to the conditions of the real experiment. The simulation area was $20\times 20$ cm, the plasma flow diameter was 2 cm, the ion concentration in the plasma flows was $10^{15}$ cm$^{-3}$, the plasma was assumed to be fully ionized and consist of singly ionized aluminum, the plasma flow velocity was $V_0 = 10^6$ cm/s, the electron and ion temperature was assumed to be initially zero. The total simulation time was about 60 $\mu$s. The external magnetic field was set based on electromagnetic calculations for the known geometry of real coils. The field strength in the bases of arch was $B_0 = 80$ mT.

With the specified parameters, the ion inertial length is about 1.2 cm, the ion Larmor radius is of the same scale at magnetic Mach numbers of about 1, and the plasma jet diameter is 2 cm. This leads to dynamics features associated with the FILR. To reveal these features in the simulations, a series of calculations were also carried out for larger scales of the system: in a $120\times 120$ cm box and for a plasma jet diameter of 12 cm.

In the first approximation, the effects of FILR can be described as an anomalous viscosity, the so-called gyroviscosity, characterized by the kinematic viscosity coefficient $\nu = V_{i\perp}^2/2\Omega_i$, where $V_{i\perp}$ is the thermal velocity of ions across the magnetic field, $\Omega_i = ZeB/M_i$ is the ion gyrofrequency. The Reynolds number can then be estimated as $R = V_iL/\nu = 2(V_i/V_{i\perp})(L/r_i)$, where $V_i$ is the ion velocity, $L$ is the scale of the inhomogeneity, $r_i$ is the ion Larmor radius. Note that in the case of weak curvature of magnetic lines, so that their radius of curvature is $\varrho_B \gg r_i$, the transverse velocity of ions changes adiabatically, and in a quasi-homogeneous field remains constant, thus, for supersonic flows $V_i/V_{i\perp} \sim 5$. However, in strongly curved fields, for which $\varrho_B \lesssim r_i$ the longitudinal velocity quickly turns into transverse, and thus, $V_i/V_{i\perp} \sim 1$.

For the first case studied (close to the conditions of the experimental setup) $\varrho_B \approx 10$ cm, and $r_i \approx 1$ cm, thus, the curvature of the magnetic lines is relatively small, although significantly higher than in the second case, for which the ion Larmor radius is the same, and $\varrho_B \approx 60$ cm. As a scale of inhomogeneity, we can take the radius of the plasma flows: $L \approx 1$ cm in the first case and $L \approx 6$ cm in the second. Thus, the Reynolds number in the first case is $R \approx 10$ and in the second case --- $R \approx 60$.

\section{\label{sec:results}Simulations results}
\begin{figure}
  \centering
  \includegraphics[width=16cm]{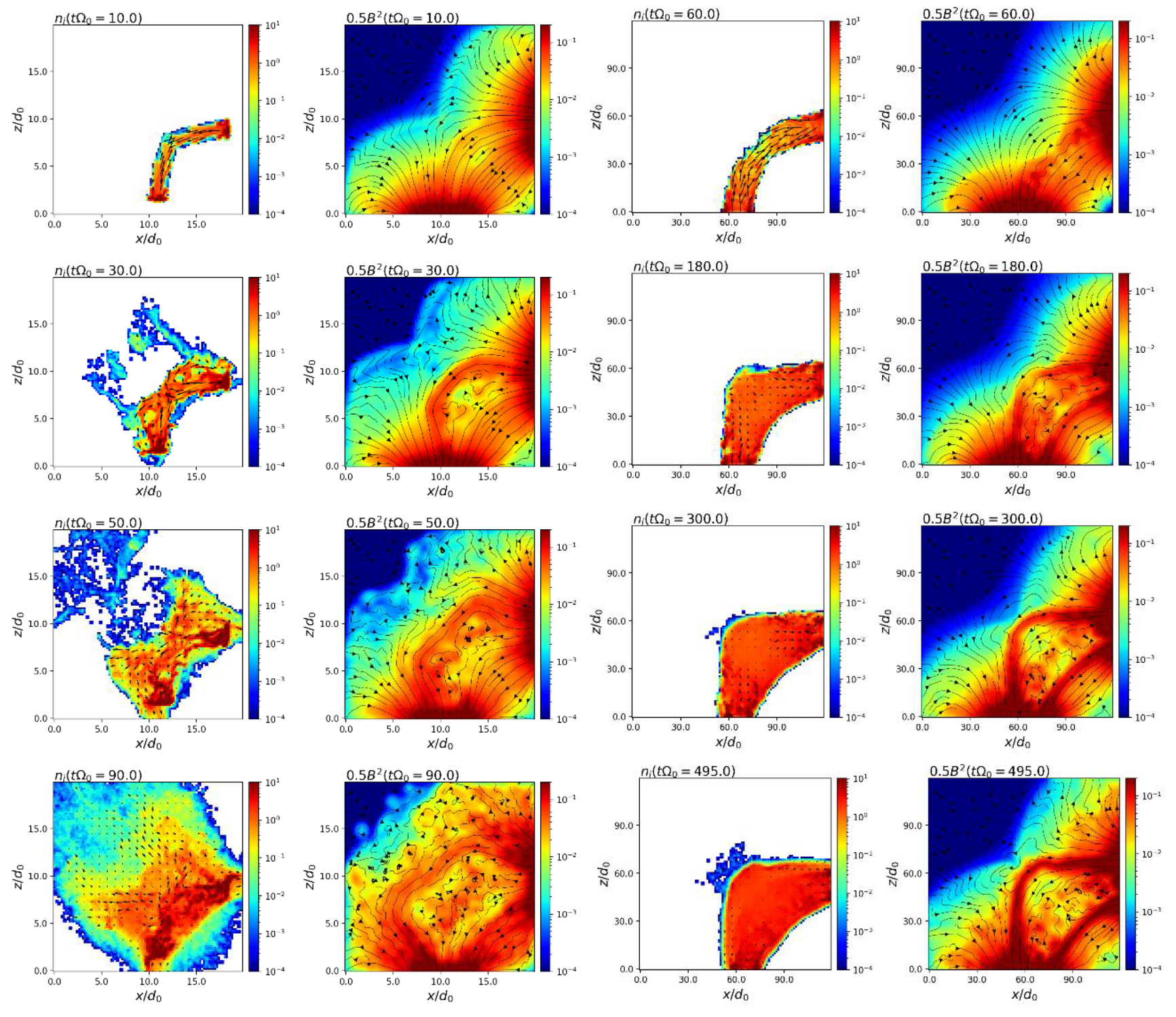}
  \caption{Ion concentration (left) and magnetic pressure (right) for calculations with small (left column) and large (right column) scale systems. Concentration is normalized to the initial concentration $N_0 = 10^{15}$ cm$^{-3}$, magnetic pressure is normalized to the initial pressure of the plasma flow ($B=1$ corresponds to 80 mT), coordinates are normalized to the ion inertial length $d_0=1.18$ cm, time is normalized to inverse ion gyrofrequency $\Omega_i^{-1} = 0.1$ $\mu$s.}
\label{fig:1}
\end{figure}

At first, the sub-Alfvenic regime was investigated, in which the magnetic Mach number was slightly less than unity (the plasma flow pressure was lower than the magnetic pressure). The evolution of the plasma density and magnetic field is shown in Fig. \ref{fig:1}.

Due to the spatial scales increased by 6 times in the second case, the given time moments are 6 times greater than the time moments in the first case. As we can see, the the interaction has noticeable differences. For large-scale flows, a calmer, quasi-stationary interaction pattern is observed. Both ions and electrons are magnetized, the general dynamics are of MHD nature. A stable plasma arch is formed, which practically does not evolve in time. For a small-scale system, the formed arch is not stationary, a part of the plasma escapes it and its fairly rapid expansion is observed. Let us estimate the time required for such expansion due to gyroviscosity. We will consider the diffusion coefficient to coincide with the kinematic viscosity coefficient: $D\approx\nu$. The expansion time can then be estimated as $T_D \approx \mathcal{L}^2/D = R(\mathcal{L}/L)(\mathcal{L}/r_i)\Omega_i$, where $\mathcal{L}$ is the distance over which the plasma has expanded. For the first case, we have $\mathcal{L} \approx 10$ cm and $T_D \approx 1000\Omega_i$, which significantly exceeds the simulation time. Thus, the plasma expansion in this case cannot be explained by the gyroviscous model alone, but occurs mainly due to the $\mathbf E\times\mathbf B$ drift.

\begin{figure}
  \centering
  \includegraphics[width=16cm]{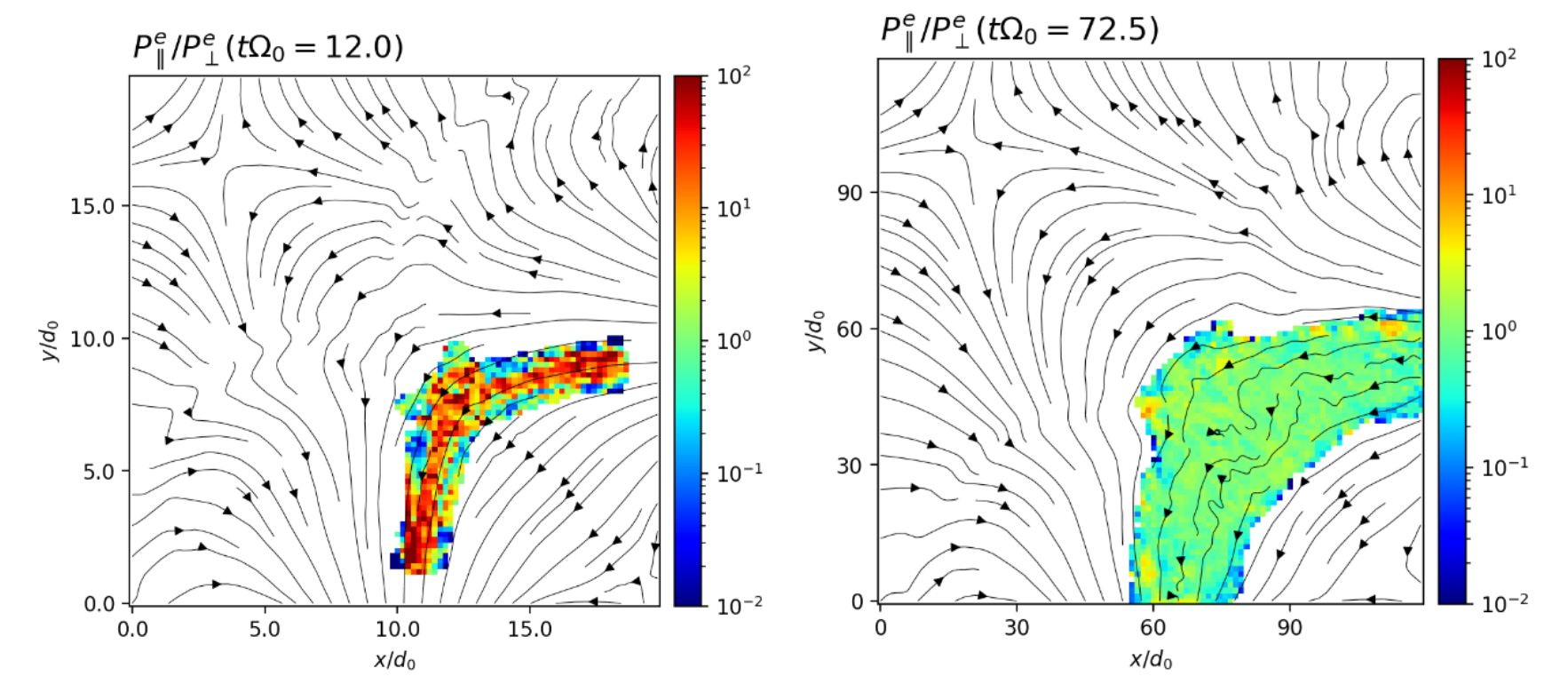}
  \caption{The ratio of the longitudinal component of the electron pressure tensor with respect to the magnetic field to the transverse component in the calculation plane. On the left is a calculation with the small scale system, on the right is a calculation with the large scale system.}
\label{fig:2}
\end{figure}

Note also that in the small-scale system, a magnetic field region with an irregular magnetic lines is formed inside the arch, which indicates the possibility of magnetic reconnection processes occurring in it. In addition, in this case, some filamentation of the plasma density is observed, which we associate with the development of Weibel-type instability due to the anisotropy of the electron pressure tensor. Indeed, from Fig. \ref{fig:2} it is evident that the anisotropy of the electron pressure in this case is significant and reaches tens, while for large-scale flows it is close to unity.

\begin{figure}
  \centering
  \includegraphics[width=16cm]{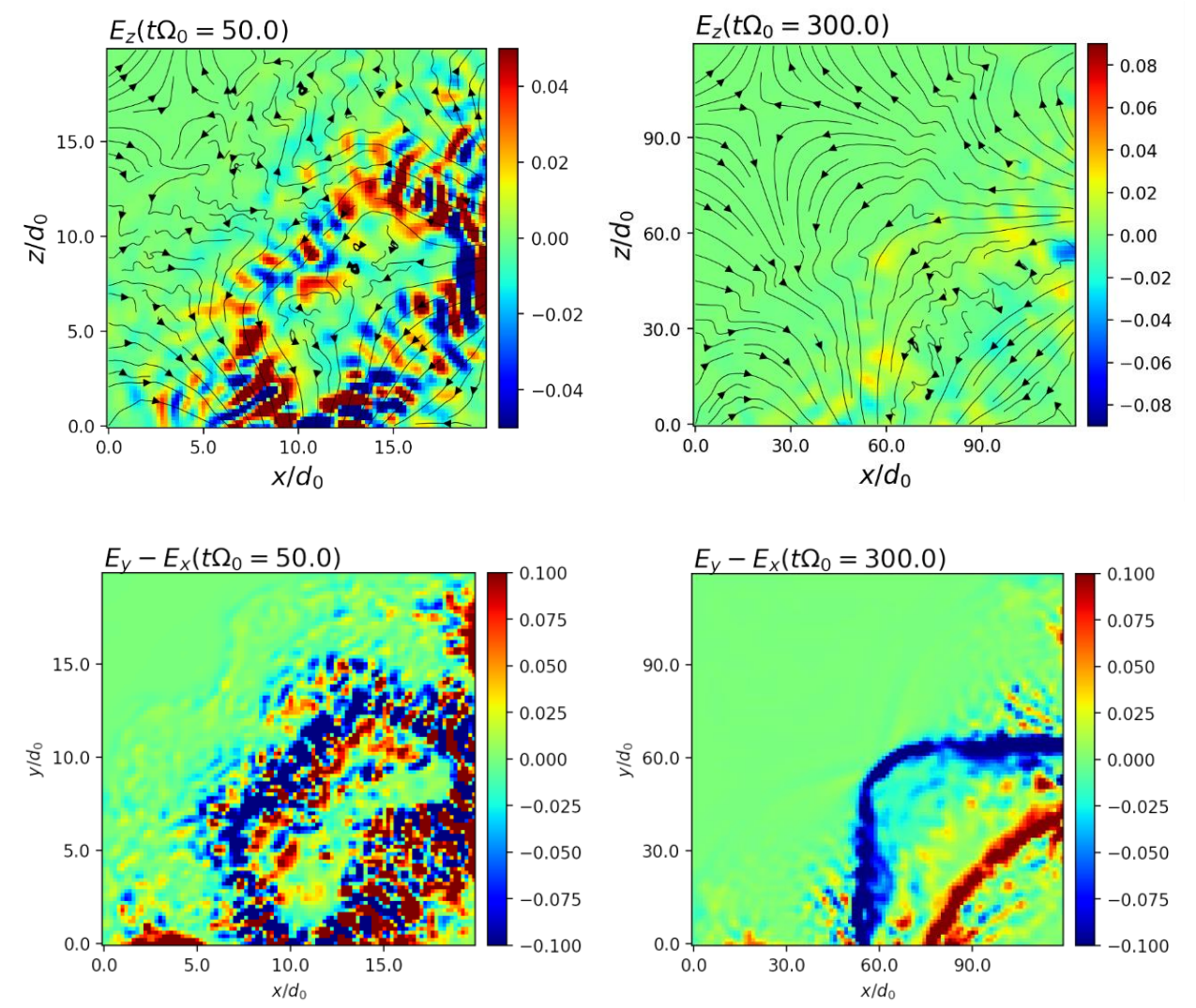}
  \caption{$z$- (top) and $xy$- (bottom) components of the electric field for small-scale (left) and large-scale (right) systems. Arrows indicate the magnetic field. The electric field is normalized to $E_0=V_0B_0=25$ V/cm.}
\label{fig:3}
\end{figure}

Another feature of the system under study is the excitation of ion-cyclotron surface waves at the plasma tube boundary. Fig. \ref{fig:3} shows that this wave is clearly seen for a small-scale system and has an elliptical polarization with a predominantly $xy$-component (i.e. this is an Alfven type wave). These waves can be excited either as a result of the development of instability associated with the anisotropy of the ion component pressure \citep{Sagdeev1961} or due to specific instabilities associated with FILR \citep{kolesnichenko_JPP_2023,kolesnichenko_PP_2024}. The later explanation is supported by the fact that for a large-scale system the wave amplitude is much weaker, and its $z$-component is not visible at all against the background of strong fields caused by the diamagnetic effect (with a spatial scale of the order of the ion Larmor radius). A more detailed study of these waves in our system, however, is beyond the scope of this paper and will be considered separately.

Thus, the small-scale system exhibits significantly less stable behaviour, and its evolution is accompanied by intensive development of instabilities of kinetic nature.

\begin{figure}
  \centering
  \includegraphics[width=16cm]{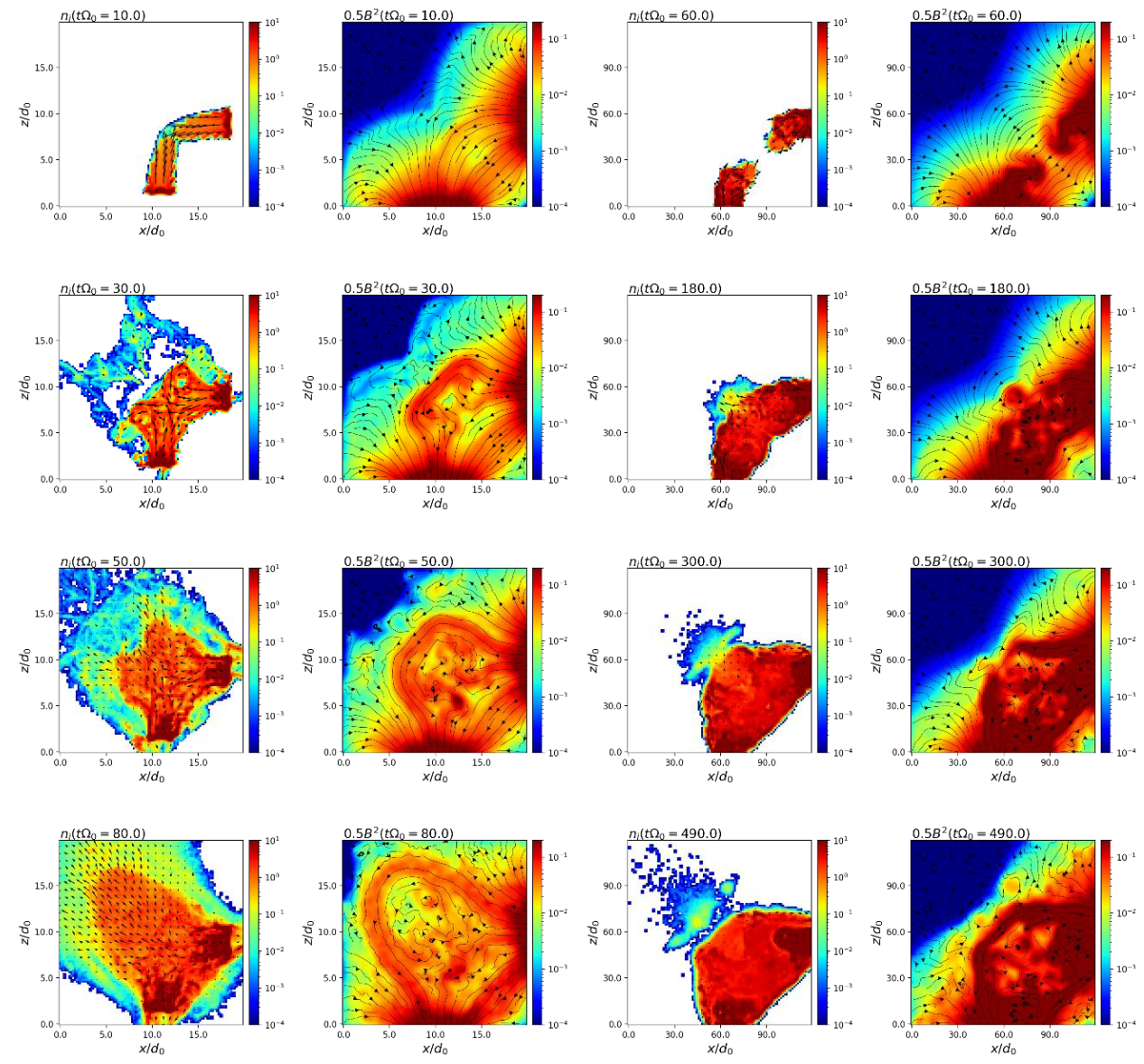}
  \caption{The same as in Fig. \ref{fig:1}, but for plasma flows 5 times denser.}
\label{fig:4}
\end{figure}

Let us now consider the case of denser flows, for which the magnetic Mach number exceeds unity (the flow pressure is higher than the magnetic pressure). The corresponding results are shown in Fig. \ref{fig:4}.

As noted in earlier studies, the interaction process in this case is noticeably more intense, and a partial rupture of magnetic lines by flows is observed. In the small-scale case, the formation of structures with closed magnetic lines – plasmoids – is observed in the expanded arch. Nevertheless, for a large-scale system, even in this case, no significant expansion of the arch is observed, although its shape is strongly deformed by the plasma pressure, and plasmoid-like structures are also observed inside the plasma.

\section{Conclusions}
Thus, the comparability of the system scales with the ion scales in the case under study is of fundamental importance for observing the intense dynamics of interaction and the development of kinetic-type instabilities. This coincides with the results of earlier studies, which noted that the FILR leads to an increase rate of existing instabilities \citep{huba_PRL_1987} and to the emergence of new ones \citep{mikhailovsky_NF_1965}. It has also recently been shown that in the arch configuration, a decrease in the arch size to the ion scale leads to the development of instabilities accompanied by intense X-ray emission of accelerated electrons \citep{zhang_NA_2023}. It allows us to expect non-thermal high frequency radiation in our experimental setup, likely, in electron cyclotron range which will be a possibly subject of our future research.

\begin{acknowledgments}
This research was funded by Russian Science Foundation grant number 23-12-00317.
\end{acknowledgments}

\bibliography{ref}

\end{document}